\documentclass[preprint,onecolumn,showpacs,preprintnumbers,amsmath,amssymb]{revtex4}

\usepackage{graphicx}
\usepackage{dcolumn}
\usepackage{bm}
\usepackage{dsfont}

\begin{document}
\preprint{}
\title{Chaos in computer performance}
\author{Hugues Berry}
\email{hugues.berry@inria.fr}
\author{Daniel Gracia P\'{e}rez}
\author{Olivier Temam}
\affiliation{INRIA Futurs, ALCHEMY,
Parc Club Orsay Universit\'{e} ZAC des vignes 4, rue Jacques Monod -
Bat G 91893 Orsay Cedex - France}
\date{\today}
\begin{abstract}
Modern computer microprocessors are composed of hundreds of millions
of transistors that interact through intricate protocols. Their
performance during program execution may be highly variable and
present aperiodic oscillations. In this paper, we apply current
nonlinear time series analysis techniques to the performances of
modern microprocessors during the execution of prototypical
programs. Our results present pieces of evidence strongly supporting
that the high variability of the performance dynamics during the
execution of several programs display low-dimensional deterministic
chaos, with sensitivity to initial conditions comparable to textbook
models. Taken together, these results show that the instantaneous performances of modern microprocessors constitute a complex (or at least complicated) system and would benefit from analysis with modern tools of nonlinear and complexity science.
\end{abstract}
\pacs{05.45.Ac,05.45.Tp, 89.20.Ff, 89.75.-k}
 \maketitle
\textbf{Modern microprocessor architectures rely on impressive
numbers of transistors (up to a billion) that interact through
intricate rules. As a consequence, the performance of these
microprocessors during the execution of certain programs displays
complex non-repetitive variations that challenge traditional
analysis. Yet, comparable complex behaviors are observed in many
other systems ranging from physics to biology and social sciences
and have been successfully described using nonlinear and chaotic
data analysis. In this paper, we apply these methods to analyze
modern microprocessor performances. We collect several measures
characterizing the architectural state and performance during the
execution of several prototypical programs and apply current
techniques of nonlinear analysis to the resulting time-varying
signals. Our results show that for several programs, the complex and
highly variable dynamics observed result from deterministic chaos.
This suggests that detailed predictions of microprocessor
performance is unlikely with these programs. Taken together, these
results show that the instantaneous performances during program
executions on modern microprocessor architectures form a byzantine
system that should benefit from analysis with modern tools of
nonlinear and complexity science.}

\section{\label{sec:intro}Introduction}
Modern computer architectures result from a rapidly growing
evolution that can be traced back to the 1960's, when Moore observed
that the number of transistors per integrated circuit displayed an
exponential growth and predicted that this trend would
continue~\cite{MooreLaw}. The so-called Moore's Law has indeed been
maintained during the last 40 years, as transistor density doubled
approximately every 18 months. Consequently, today computer
processors rely on amazingly high numbers of transistors: the
widespread Intel\textregistered~ Pentium\textregistered~ 4 contains
42 million transistors but the more recent Itanium\textregistered~ 2
possesses 410 million of them. Furthermore, a constant of this
evolution is that processor speed (especially, its clock rate) by
far outperforms memory operations. Hence, most recent advances in
the field have mainly aimed at hiding memory latencies using
engineering solutions (parallel executions,
pipelining, cache memory systems). But this necessarily came with further increases of the processor complexity.\\
As a consequence, predicting the precise performance of microprocessors (the number of instructions executed each second) during execution of programs
running on modern computer architectures has become increasingly
difficult. For instance, one efficient way to optimize computer
performance for a given program consists in fine-tuning the compiler
options to adapt the compiler work to the considered architecture.
Yet the complexity of modern architectures is such that rational
optimizations, guided by a thorough knowledge of the architecture,
are now less efficient, up to the point that more systematic
automated search methods based on
machine-learning~\cite{Stephenson2005}, genetic
algorithms~\cite{Kulkarni2004} or iterative trial-and-error
techniques~\cite{Fursin2002} are being investigated as possible replacements.\\
Hence, on the basis of the high number of their components dedicated
to performance improvement and the intricacy of their interactions,
the instantaneous performance of modern microprocessors may be
viewed as a complex system. As a consequence, performance recordings
during the execution of certain programs can be highly
variable~\cite{Duesterwald2003} and difficult to
predict~\cite{Annavaram2004}. Analyzing and predicting performance
(i.e. the rate at which the microprocessor will execute a
given program) has proven increasingly difficult.\\
Early on, computer architects dismissed modeling as inappropriate
because it was too inaccurate to capture the slight performance
differences between two architecture mechanisms.   For instance,
even modeling of a single non-trivial architecture component such as
a cache memory spawned decades of
research~\cite{377833,106981,tss,1062248}, and proved only partially
successful a few years ago for a range of programs with fairly
regular behavior and simplistic
architectures~\cite{abella01nearoptimal}. Instead, computer
architects have always relied upon detailed simulators which
describe the architecture behavior on every
cycle~\cite{simplescalar}. As a consequence, simulators execute a
program about 10000 times slower than on a real architecture, and
this technique is now becoming overly time-consuming and
inappropriate for complex processors and future processors with a
large number of cores. Consequently, novel approaches to
understanding and anticipating system behavior are
currently sought in the computer architecture community~\cite{1006729}. \\
In the present paper, we study the time-evolution of the performance during execution of several prototypical programs on
prototypical modern microprocessors. We record several
metrics characterizing execution performance (number of instruction
executed at each processor cycle) and memory operations (cache
misses). Treating these traces as
time-varying signals, we analyze them using current techniques from nonlinear time series
analysis. Besides regular periodic behaviors, we evidence highly
variable performance evolutions for several programs. More
interestingly, we show that, although the high variability displayed by
several programs can be attributed to stochastic-like sources, the
evolution of performance during the execution of several others
displays clear evidences of deterministic chaos, with sensitivities
to initial conditions that are comparable to textbook chaotic
systems.\\
The remaining of the paper is organized as follows.
Section~\ref{sec:trace} describes the setup and methodologies used
to obtain the time series we analyzed. Because of the
interdisciplinary relevance of this work and considering that we
applied a variety of methods, we present in
section~\ref{sec:timeseries} a rapid overview of the time series
analysis techniques we employed. Section~\ref{sec:bzip2} illustrates
the existence of chaotic performance trace with the example of the
execution of the program \verb"bzip2". Stochastic-like performances
are also evidenced in section~\ref{sec:vpr} and the example of the
program \verb"vpr". Finally, we present for comparison in
section~\ref{sec:applu} the performance displayed during
\verb"applu" execution, as a prototype of regular periodic
evolution. Section~\ref{sec:Conclusion} discusses possible
explanations for the observed behaviors and present potential
implications in practical applications.
\section{\label{sec:trace}Program traces}
The time series shown in this article were obtained using a
processor \emph{simulator}.  A simulator is a large program that
implements a detailed description of the computer microarchitecture
(at the level of a clock cycle and bits), and it is the tool used by
computer architects to design and try out various processor options.
The simulator is fed with an instruction trace, corresponding to a
given program executing a given data set.  And the purpose of the
simulator is to understand how many cycles are necessary to execute
this instruction trace, as well as to expose the internal operations
of the processor for analysis.\\
A real processor, such as the Pentium 4, also embeds hardware
counters that collect some statistics on its internal operations.
However statistics are sampled infrequently  (and thus too coarsely)
in order to avoid disrupting normal processor operations,  which is
not appropriate in our case.  Also, such counters make it hard to
distinguish between the multiple programs (and the operating system)
which time-share the processor,  so that it is not obvious or just
impossible to reconstruct the time series for a single program.\\
Still, the simulator we used, called
SimpleScalar~\cite{simplescalar}, corresponds to the architecture of
a typical modern superscalar processor (the Pentium 4 is also a
superscalar processor). It is currently used in more than 50\% of
computer architecture research articles. It has been validated at
15\% accuracy against a fairly recent superscalar processor (the HP
Alpha 21264) used in many servers~\cite{1054909}.\\
On this simulator, we ran the 26 Spec benchmark programs composing
the so-called Spec suite (we used the Spec2000 version of the
benchmark suite). A benchmark is a program selected as
``representative'' of an application domain. And the Spec benchmark
suite is the most widely used to evaluate and compare the
performance of new computer and processor architectures. Each
benchmark comes with three data sets, with two data sets being
voluntarily small and medium size (respectively labeled
\texttt{test} and \texttt{train}).  All experiments in this article
were conducted with the third and most realistic data set, called
\texttt{ref} (for ``reference''). In some cases (e.g. \verb"bzip2"),
the \texttt{ref} data set proposes
several input data.\\
During the execution, we collected 3 performance metrics:  the IPC,
the L1 and L2 miss rates. The IPC stands for the average number of
\emph{Instructions Per Cycle} and is the typical global performance
metric for superscalar processors.  L1 and L2 respectively
correspond to the first-level and second-level cache, small and fast
memories used in all processors and aiming at hiding the main memory
latency.  The L1 and L2 form a memory hierarchy, with the L1 being
closer to the processor, and smaller but faster than the L2. The
miss rate is the percentage of processor requests that cannot be
served by the cache (the request is then sent to the lower level of
the hierarchy), and it thus characterizes the cache efficiency. The
cache behavior has a strong impact on performance, so besides the
global IPC metric, the caches miss rates are key performance
metrics.\\
Running an entire program requires the execution of several billion
instructions, so that it is technically impossible to handle
execution traces that would both cover the entire program execution
and display the value of the chosen metric for \textit{each clock
cycle}. Furthermore, modern microprocessors rely heavily on
speculative execution: upon encounter of a conditional branching,
the microprocessor begins to execute one of the branch alternative
before the outcome of the conditional branch test is known (i.e.
before the microprocessor knows which branch should actually be
taken). In other words, at a given clock cycle, the microprocessor
might be executing several instructions that can possibly be
discarded from the program flow a moment later. In this framework,
measuring performance is meaningful only if measurements are time
averages. Accordingly, our execution traces present \textit{averages
of the metric} over a certain number $\tau_{av}$ of consecutively
executed instructions (where we have used $\tau_{av}=10^6,\,10^7\,
\mathrm{or}\,10^8$ instructions).
\section{\label{sec:timeseries}Time series methods}
Nonlinear time series methods are based around dynamical systems
(continuous-time ordinary differential equations and iterated maps).
Hence, they can be powerful tools for analyzing microprocessor
behaviors only if they display the same computation power as
microprocessors. More specifically, because microprocessors are
capable of universal computing (they are Turing machines), they
should also be universal. Recent works have clearly stated that
dynamical systems are indeed capable of universal computation. For
instance, discrete-time dynamical systems are computationally
universal, as several of them have been demonstrated to be able to
simulate the computation of a Turing machine. This is the case of
piecewise-linear maps in $\mathds{R}^2$~\cite{Koiran1994}, cellular
automata~\cite{Cook2004}, and neural networks (especially recurrent
networks with rational or real weights and saturated
linear~\cite{Kilian1993} or sigmoid~\cite{Siegelmann1991} activation
function). Universal computation has also been evidenced for several
continuous-time dynamical systems, including ordinary differential
equations~\cite{Branicky1995}, partial differential
equations~\cite{Omohundro1984}, and continuous-time Hopfield neural
networks~\cite{Sima2003}. Hence, analysis techniques based on
dynamical systems, such as nonlinear time series methods, are
susceptible to be
powerful tools for analyzing microprocessor behaviors.\\
The program
traces were analyzed using a variety of methods for nonlinear time
series analysis that we briefly present in this section. Note that
for most of these methods, we used the TISEAN routine
package~\cite{tisean,Schreiber1999}.\\
Let $\left\lbrace x(1), x(2), x(3),\ldots x(N)\right\rbrace $ be the
time series under consideration. Each value $x(n)$ of the time
series is the average of the metric over a number $\tau_{av}$ of
consecutively executed instructions (see~\ref{sec:timeseries}). In
other words, $x(n)$ represents the average value of the metric
between the execution of instruction number $n\tau_{av}$ and that of
instruction number $(n+1)\tau_{av}$. For this reason, we can
reasonably consider that the state-space of our time series is
continuous. Accordingly, the continuous nature of our measurements
can readily be judged from visual inspection of these time series.
Indeed, in every figure of the paper, we plot the obtained values as
isolated dots, i.e. \textit{we do not join successive values with
lines}. Hence, the continuous aspect of the curves plotted on
Figure~\ref{fig:1} A \& B, for instance, is not a plotting artifact,
but reflects the continuity of the values adopted by the successive
values of the time series.
\subsection{\label{sec:Spectral}Temporal correlations}
To study the presence of temporal correlations amongst time series,
we used two complementary methods: spectral analysis and detrended
fluctuation analysis. Spectral analysis is based on the Fourier
spectrum of the time series. If a sequence has long-range (power-law) correlations, its power spectrum
$S(f)$ is related to the frequency $f$ through a power law
\begin{equation}
S(f) \propto f^{-\beta}
\end{equation}
where $\beta$ is the spectral exponent. Uncorrelated white noise
contains all possible frequencies and is characterized by the
exponent $\beta = 0$. So called "fractal" time series display
strictly positive $\beta$. For instance, $1/f$-noise defines signals
with $\beta \approx 1$ while $\beta = 2$ for Brown
noise~\cite{Yamamoto1999}.
\\Contrarily to spectral analysis, detrended fluctuation analysis (DFA) permits the detection of
long-range correlations in nonstationary data (i.e. signals
that do not display a constant mean value) and avoids spurious detections of apparent
long-range correlations that are possible with spectral
analysis~\cite{Peng1995}. The time series is first integrated: $y(k)=\sum_{i=1}^k
\left[x(i)-\overline{x}\right]$, where $x(i)$ is the $i$th value of
the time series and  $\overline{x}$ its average over the series. The
integrated time series is then divided into time windows of equal
duration $n$. In each window, the least-squares fitted line (the
local trend) is computed. The $y$ coordinate of the straight line
segments is denoted by $y_n(k)$. The integrated signal $y(k)$ is
next detrended by subtracting the local linear trend $y_n(k)$ in
each window. The average root-mean-square fluctuation of this
integrated and detrended time series is computed as
\begin{equation}
F(n)=\sqrt{\frac{1}{N}\sum_{i=1}^N[y(i)-y_n(i)]^2}
\end{equation}
The procedure is repeated over all time scales (window duration)
$n$. Typically for fractal time series, $F(n)$ increases as a
power-law of $n$
\begin{equation}
F(n) \propto n^{\alpha}
\end{equation}
A value of $\alpha = 0.5$ characterizes an uncorrelated signal, such
as a white noise, whereas $\alpha>0.5$ indicates the presence of
long-range positive (persistent) temporal correlations. Note that
periodic signals have $\alpha = 0$ for time scales larger than their
period of repetition.
\\These tests are complementary because it has been evidenced that, using one of these
methods alone, the presence of long-range correlations may be
artifactually detected, while agreement between independently
obtained values of $\alpha$ and $\beta$ according to theoretically
derived relationships limits the risk of spurious
determinations~\cite{Rangarajan2000}.
\subsection{\label{sec:Embedding}Embedding}
Most dynamical systems possess many degrees of freedom and take
place in multi-dimensional phase space. Yet, the vast majority of
real-life time series are single-valued, and even if multiple
simultaneous measurements are available, they rarely are in
sufficient number to cover all the degrees of freedom of the system.
However the missing information can be recovered by reconstructing
the original attractor on the basis of a single-valued time series.
Actually, the evolution of any single variable of a dynamical system
is determined by the other variables with which it interacts. The
basic idea of embedding methods for attractor reconstruction is thus
that information about the relevant variables is implicitly
contained in the history of any single variable. A delay
reconstruction with delay time $\tau$ and embedding dimension $m$ is
obtained by forming a new vector time series ${\mathbf{X}(t)}$ in an
$m$-dimensional embedding space according to
\begin{equation}
    \mathbf{X}(t)=\left(x(t),x(t+\tau),\ldots,x(t+(m-1)\tau)\right)
\end{equation}
Takens' embedding theorem~\cite{Takens1981} states that, for
sufficiently large $m$, the geometry of ${\mathbf{X}(t)}$ in the
embedding space captures the topological properties of the original
attractor in its natural phase-space. Hence, characterization
methods originally dedicated to the original attractor can identically be applied to the reconstructed one~\cite{Packard1980}.\\
The determination of "optimal" values for the embedding parameters
is a delicate step in attractor reconstruction because this
procedure can amplify noise in real-life time
series~\cite{Casdagli1991}. There are currently two major methods
for estimating the time delay $\tau$. The first consists in setting
$\tau$ as the time necessary to cancel the correlation between two
time series values and thus selecting the first zero-crossing of the
signal auto-correlation function or the time at which it has dropped
to $1-1/e$ of
its initial value~\cite{Rosenstein1993}. An alternative approach sets $\tau$ as the first minimum of the time delayed (average) mutual
information function~\cite{tisean}. The question of which of these two methods should be used is still an open
problem~\cite{Abarbanel1996, Kantz1996}. In this paper, we estimated for each data sets both the first zero-crossing of the
autocorrelation function and the first minimum of the average mutual information. In the rare cases where the corresponding estimates
were not similar, we set $\tau$ to the value given by the latter method.\\
The most frequent method for determining the embedding dimension $m$
is called the \textit{false nearest neighbor} method~\cite{tisean}.
Briefly, suppose the correct embedding dimension is $m_0$, i.e. for
$m=m_0$, the reconstructed attractor is a one-to-one image of the
original one. If one attempts to embed the time series in a
$m$-dimensional space with $m<m_0$, the topology of the attractor
will not be conserved, so that several points will be projected into
neighborhoods of other points, to which they would not belong in
higher dimensions. Hence, if two points are found in proximity in
the embedding space, this can be due either to the dynamics that
brought them close, or to an overlap resulting from the projection
of the attractor to an insufficient dimension, in which case these
points are referred to as `false neighbors'. By comparing the
Euclidean distance between two points in consecutive embedding
dimensions $m$ and $m+1$, it is possible to quantify the percentage
of false neighbors at embedding dimension $m$~\cite{Kennel1992}. The
optimal dimension is then defined as the minimal dimension for which
the percentage of false neighbors is zero or at least, sufficiently
small.
\subsection{\label{sec:Rec}Recurrence plot}
Recurrence plots are graphical representations suited to qualitatively assess the presence of patterns and nonlinearities, even in short and nonstationary time series~\cite{Eckmann1987}. It consists in computing the distances between all pairs of vectors in the embedded time series, applying a threshold $\xi$ to the resulting distance matrix
\begin{equation}\label{Eq:RP}
\mathbf{R}_{i,j}=\Theta
\left( \xi-\parallel \mathbf{X}(i) -
\mathbf{X}(j)\parallel\right)\qquad i,j=1,\ldots,p
\end{equation}
where $p$ is the number of points of the attractor, $\Theta(\cdots)$
is the Heaviside threshold function:
\begin{equation}\nonumber
\Theta(x)=\left\{\begin{array}{ll}1&x\geq0\\0&x<0\end{array}\right.
\end{equation}
 and $\parallel\cdots\parallel$ denotes 2-norm. Recurrence plots are two-dimensional graphical representations of this thresholded distance matrix that assign "black" dots to the value one, and "white" dots to the zero value. The value of the threshold $\xi$ was estimated according to Zbilut \textit{et al.}, 2002~\cite{Zbilut2002}. In the case of a deterministic signal, whenever a point $\mathbf{X}(i)$ is found close to another point $\mathbf{X}(j)$ in the embedding space, then the points $\mathbf{X}(i+1), \mathbf{X}(i+2), \ldots, \mathbf{X}(i+k)$ will likely be close to $\mathbf{X}(j+1), \mathbf{X}(j+2), \ldots,\mathbf{X}(j+k)$. Hence, deterministic signals are characterized by recurrence plots with black diagonal lines parallel to the minor diagonal. Alternatively, stochastic processes manifest as single isolated black points forming more homogeneous and random patterns. Chaotic signals are deterministic systems with high sensitivity to initial conditions (see below). Accordingly, their recurrence plots are characterized by broken diagonal lines beside single isolated points. Plots with fading to the upper left and lower right corner usually indicate a drift, i.e. nonstationarity in the time series.
\subsection{\label{sec:Poinc}Poincar\'{e} sections}
The goal of Poincar\'e section is also to detect structures in the
attractor. It consists in building $m-1$-dimensional cross-sections
transverse to the $m$-dimensional attractor and collecting the
corresponding successive intersections according to one direction
(crossing from the ``bottom'' side to the ``top'' side for example).
The corresponding Poincar\'e map (or first-return map) is obtained
as a plot of each intersection as a function of the next one.
Alternatively, it is possible to define the cross-section surface by
the zero crossing of the temporal derivative of the signal, thus
collecting maxima or minima~\cite{tisean}. In the present paper,
Poincar\'e maps were constructed using minima. Roughly speaking,
Poincar\'e maps of stochastic systems show homogeneously distributed
and space filling patterns while deterministic components form
extended low-dimensional structures.
\subsection{\label{sec:Correl}Correlation dimension and entropy}
Chaotic trajectories in dissipative systems must overcome two opposite constraints in the phase space. In the one hand, dissipation contracts volume elements under the action of the dynamics, so that the distance between two neighbors in the phase space must globally diminish with the dynamics. On the other hand, these systems display a high sensitivity to initial conditions (see below), meaning that two neighbor trajectories in the phase space diverge exponentially with time (at least locally). Hence, to accommodate these two constraints, most strange attractors present a heavily folded and complex structure, which is very often self-similar and fractal. The correlation dimension $D_2$ is one measure of the attractor fractality and is usually determined by computing the correlation sum. Briefly, it consists in determining the average probability to find two data points belonging to the attractor in a neighborhood of size $\epsilon$ in the $m$-dimensional embedding space
\begin{equation}\label{Eq:CorSum}
C(m,\epsilon)=\frac{2}{p(p-1)}\sum_{i=1}^n \sum_{j > i}^n\Theta
\left( \epsilon-\parallel \mathbf{X}(i) -
\mathbf{X}(j)\parallel\right)
\end{equation}
Note the similarity with the definition of the recurrence plots (Eq.~\ref{Eq:RP}). Indeed, estimation of correlation dimension and entropy on the basis of recurrence plots has recently been proposed~\cite{Thiel2004}.\\
If the time series is characterized by an (possibly strange) attractor, then for sufficiently small $\epsilon$ values and when $m>D_2$
\begin{equation}\label{Eq:CorSumScaling}
C(m,\epsilon)\approx \mathrm{e}^{-m h_2} \epsilon^{D_2}
\end{equation}
Alternatively, stochastic systems form trajectories that uniformly fill the $m$-dimensional embedding space so that in this case, the
correlation sum is expected to scale with the embedding dimension $C(m,\epsilon)\propto \epsilon^m$. Hence, log-log representations of
 the correlation sums $C$ against $\epsilon$ for increasing $m$ values should display linear zones with saturating slopes at high $m$
 (scaling region) in the case of chaotic dynamics, or increasingly large ones in the case of stochastic dynamics. A more accurate way to
 detect these scaling regions is to estimate the corresponding local slopes given by $d \ln C(m,\epsilon)/d \ln \epsilon$ and plot them
 against the corresponding $\epsilon$ values~\cite{tisean}. In the case of chaotic dynamics, the corresponding curves at various $m$
 should collapse onto an $m$ and $\epsilon$-independent behavior (in the scaling regions) that directly yields $D_2$. Such a collapse is
 not observed with stochastic signals. Note that an important precaution in computing the correlation sums is to exclude temporally
 correlated points from the pair counting in eq.\ref{Eq:CorSum}~\cite{Theiler1990} by ignoring all pairs of points with time indices
 differing by less than $w$ (the so-called Theiller windows $w$). In this paper, we have used $w=20$ million instructions. \\
Another quantifier of the attractor is the correlation (order-2
R\'eny) entropy $h_2$, which is obtained through the $m$-dependence
of Eq~\ref{Eq:CorSumScaling} inside the scaling regime. The
correlation entropy is usually considered as a lower bound of the
sum of the positive Lyapunov exponents~\cite{tisean}.
\subsection{\label{sec:lambdamax} Largest Lyapunov exponent}
Sensitivity to initial conditions is a hallmark of chaotic systems.
Its implies that two trajectories found in an arbitrary small
neighborhood of the phase (or embedding) space diverge exponentially
with time, thus abolishing predictability in these systems. Consider
two neighbor points $\mathbf{X}(i)$ and $\mathbf{X}(j)$ in the
embedding space and denote their distance $\delta_0 = \parallel
\mathbf{X}(i) - \mathbf{X}(j)\parallel$. After a time $t$, their
distance $\delta_t$ is expected to grow exponentially
\begin{equation}
\delta_t=\parallel \mathbf{X}(i+t) - \mathbf{X}(j+t)\parallel \approx \delta_0 \mathrm{e}^{\lambda_{max} t}
\end{equation}
where $\lambda_{max}$ is the largest Lyapunov exponent. In general,
in a $m$-dimensional space, the rate of expansion and contraction of
the  trajectories is described for each dimension by a different
Lyapunov exponent. However, estimation of the largest one is both
much easier to compute than the whole spectrum and sufficient to
decide about the presence of deterministic chaos in the data (i.e.
the largest Lyapunov exponent is expected to quickly dominate the
distance growth). To estimate $\lambda_{max}$, Kantz's
method~\cite{Kantz1994} consists in selecting a point
$\mathbf{X}(i)$ and searching all the points $\mathbf{X}(j)$ present
in a neighborhood $\mathcal{U}_i$ of $\mathbf{X}(i)$. One then
computes the average quantity $S$ (stretching factor)
\begin{equation}\label{Eq:Slmax}
S\left( \epsilon,m,t\right)=\left\langle \ln\left(
\frac{1}{p_i}\sum_{\mathbf{X}(j)\in \mathcal{U}_i} \parallel
\mathbf{X}(i+t) - \mathbf{X}(j+t)\parallel \right)  \right\rangle
\end{equation}
where $p_i$ is the number of points in $\mathcal{U}_i$ and
$\epsilon$ its size, and $\left\langle \right\rangle$ indicates
averaging over all the points in the time series. In the case of
chaotic dynamics, a plot of $S\left( \epsilon,m,t\right)$ against
time $t$ will yield a linear increase at short times for a
reasonable range of $\epsilon$ and sufficiently large $m$. The slope
of this linear regime can be used as an estimate of the largest
Lyapunov exponent $\lambda_{max}$. An alternative method, proposed
by Rosenstein~\cite{Rosenstein1993}, only considers the closest
point $\mathbf{X}(j)$ of each reference point $\mathbf{X}(i)$ in
Eq~\ref{Eq:Slmax}.
\subsection{\label{sec:surr} Surrogate data testing}
Surrogate data testing is a method to statistically infer the presence of nonlinear processes in time series. The idea is to generate artificial linear time series (surrogates) with the same power spectrum, the same correlations, and the same
distribution of values than the series to be tested~\cite{Surrogates}. The two time series are then characterized by a statistics that quantifies nonlinearity in time series with a single number. In the present work, we have used two statistics: a nonlinear (locally constant) predictor
error statistics and a time-reversal asymmetry (third order) one~\cite{Surrogates}. These results are then used to perform a statistical test in which the null hypothesis states that the series to be tested could be generated by a linear process such as that used to generate the surrogate~\cite{Surrogates}.
\begin{figure*}[!htb]
  \includegraphics[scale=0.7]{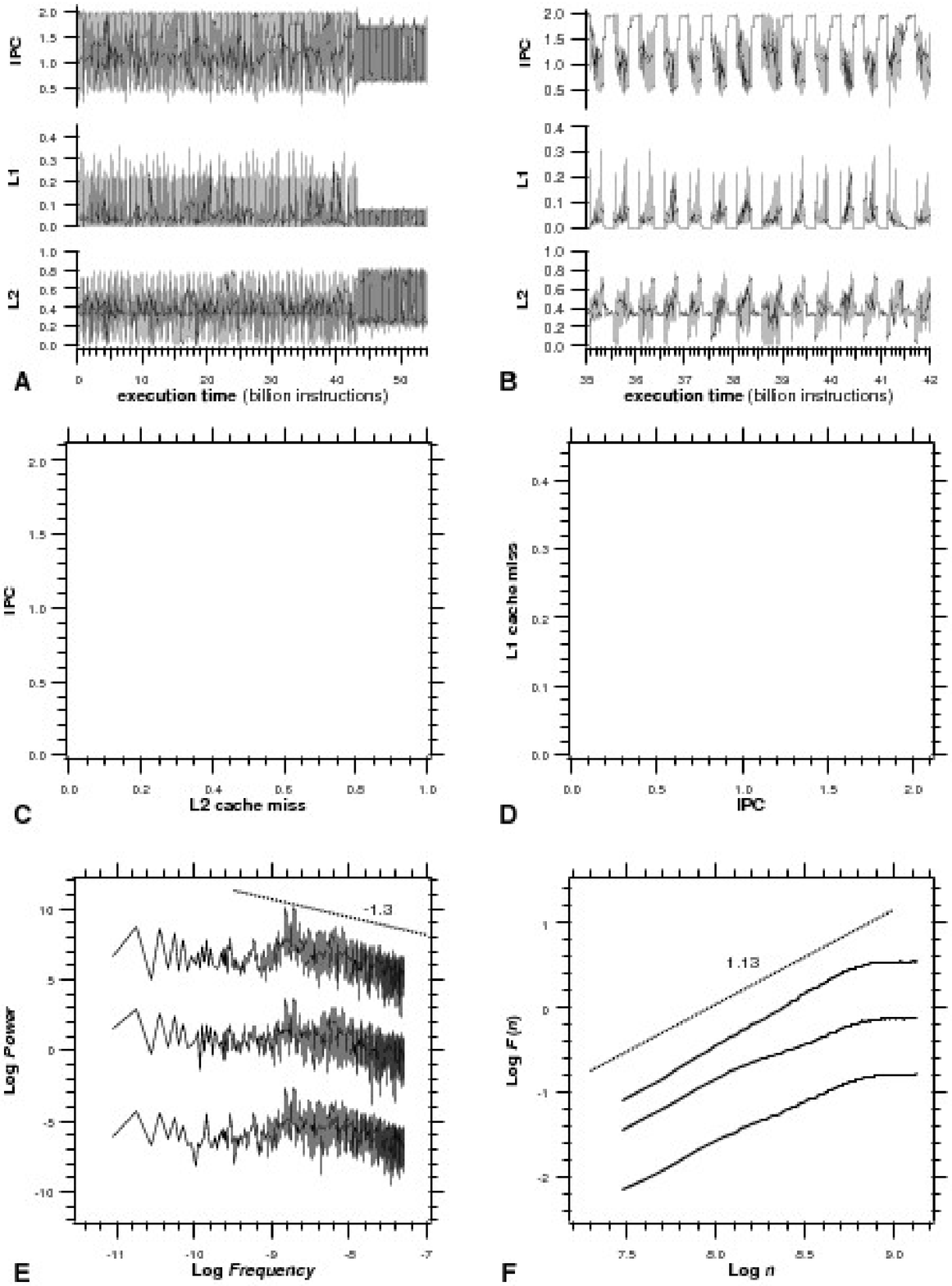}
  \caption{\label{fig:1} Execution traces for the program bzip2.
  (A) Evolution of the number of instructions executed per cycle (IPC), the rate of L1 cache misses (L1)
  and that of L2 cache misses (L2) during the first 54 billion instructions and (B) enlargement of the part comprised between 35
  and 42 billion instructions. Also shown are projections of the dynamics attractor (C) in the IPC-L2 cache miss rate phase
  plan and (D) in the L1-IPC phase plan. The presence of long term correlations in these signals was sought for by
  Power Spectrum and Detrended Fluctuation Analysis (DFA). The power spectrum is presented in (E) as a log-log plot of the data
  concerning, from bottom to top, L1, L2 and IPC traces, respectively. The curves have been arbitrarily shifted along the y-axis
  to avoid overlap. Frequency is expressed in $instruction^{-1}$. The dashed line indicates a power-law decrease with exponent 1.3.
  The DFA graph (F) displays on a log-log scale the detrended fluctuations $F(n)$ as a function of the time scale $n$
  (in instructions) for, from bottom to top, L1, L2 and IPC traces, respectively. Here, the dashed line indicate power-law growth with
  exponent 1.13. Note that the three DFA curves flatten when $n$ becomes higher than the major repetition period ($\approx 0.6$ billion
  instructions, see B).}
\end{figure*}
\section{\label{sec:res}Results}
\subsection{\label{sec:bzip2} First example: bzip2 time series}
Figure~\ref{fig:1} displays performance statistics for the program
\verb"bzip2" acting on the \verb"source" input of the \verb"ref"
data set (see~\ref{sec:trace}). We focus here on three statistics
that are particularly relevant to computer performance: the number
of instruction executed at each computer cycle (IPC), the
instantaneous rate of L1 cache miss rate (L1) and that concerning L2
cache (L2). For readability, we only display in Figure~\ref{fig:1}A
the traces obtained for the first 54 billion executed instructions
(i.e. approximately one half of the total program execution). The
three traces show two distinct phases: a first one with higher
variability and lower frequency (up to circa 43 billion
instructions), followed by a phase characterized by higher frequency
and lower variability (from 43 to 54 billion instructions). Note
that the second part of the total execution trace (not presented in
Figure~\ref{fig:1}) essentially consists of a repetition of these
two consecutive phases. In the remaining of this section we treat
the entire ($\approx110$ billion instruction long) trace as a single
entity. Note however that we have also studied the two \verb"bzip2"
execution phases separately (i.e. restraining the time series to the
first phase, from 1 to 50 billion instructions, or to the second
one, from 50 to 54 billion instructions) and obtained qualitatively
similar results (though sensitivity to initial conditions seems
higher in the second phase).\\
\begin{figure*}[!htb]
  \includegraphics[scale=0.7]{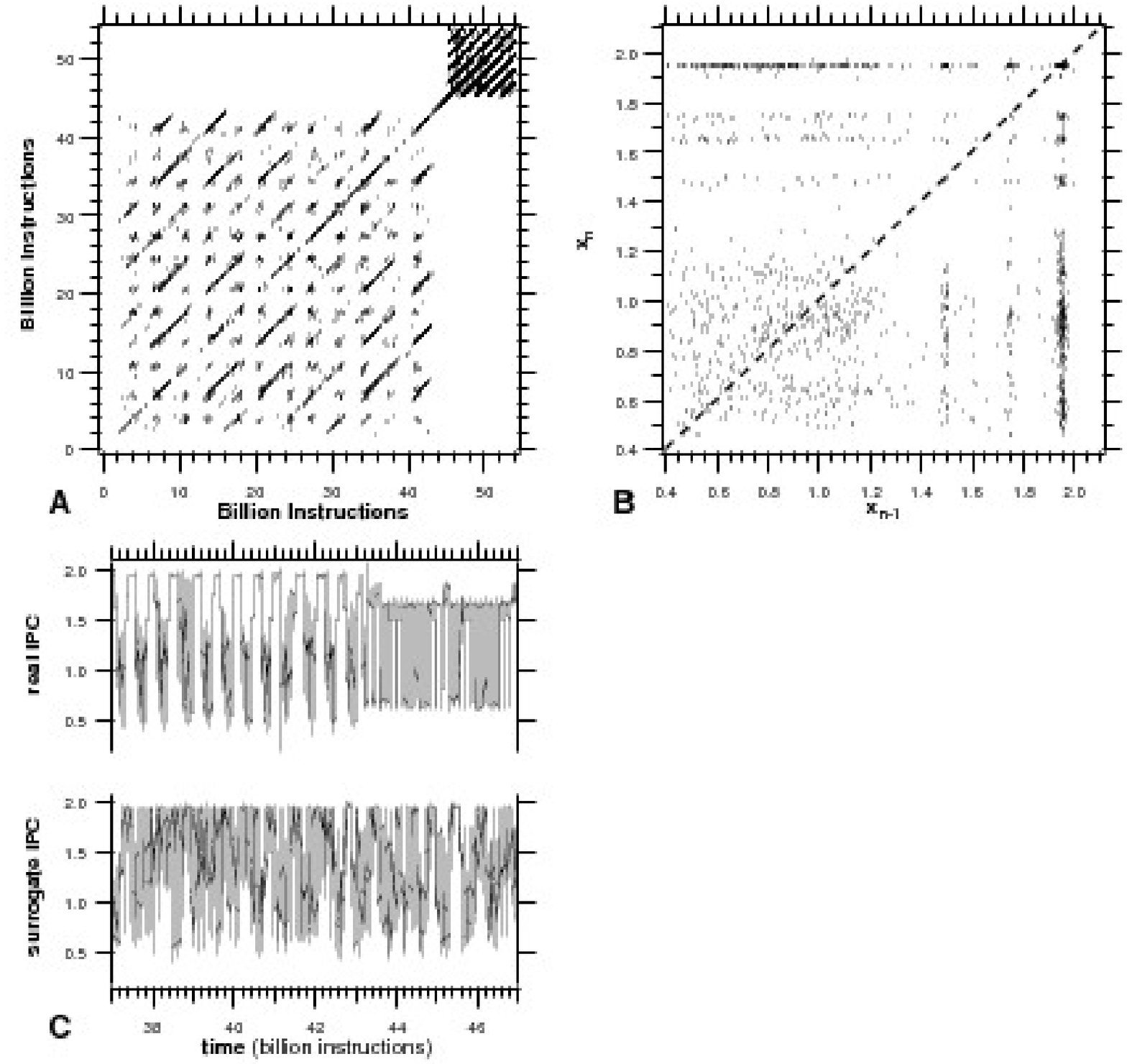}
  \caption{\label{fig:2} Study of the IPC trace for bzip2 after attractor reconstruction by embedding. (A) Recurrence plot corresponding to Figure~\ref{fig:1}A (thresholded at
  0.136) (B) First return map for the Poincar\'e section at IPC minima. The dashed line indicates the
  first diagonal. (C) Surrogates data generated from the complete IPC trace. Shown are enlargements of the part comprised between 37
  and 47 billion instructions of the original IPC time series (upper trace) and the corresponding surrogate trace (lower trace). Embedding parameters: delay
  $d=153$ million instructions, dimension $m=14$.}
\end{figure*}
Although some regularity is readily seen in these time series, the
two phases clearly display irregular or noisy dynamics. This is
especially visible from the enlargement displayed
Figure~\ref{fig:1}B. The dynamics present bounded and somewhat
regular variations together with a large amount of variability. In
particular, this figure evidences a major period of repetition of
$\approx 0.6 \times 10^{9}$ instructions. Figure~\ref{fig:1}C and D
show projections of these dynamics in the IPC-L2 and L1-IPC phase
plans. The resulting attractor projections display a characteristic
mixture of regular structured zones together with "cloudy" areas,
hence confirming the high variability of the time series. \\The
observed variability could be imputed to a noise source (as part of
the dynamics itself or resulting from the sampling method).
Alternatively, it could be a direct result of deterministic chaotic
dynamics. To discriminate between both possibilities, several tests
are available in the time series analysis literature. These tests
are usually individually conclusive when employed on long and
perfect synthetic time series. Real world time series however,
usually incorporate high levels of noise stemming from experimental
measurements, and are often much smaller, so that conclusive
decisions generally need the investigation of the results provided
by several of these tests. Thus, several converging approaches are
necessary to identify nonlinear patterns and avoid spurious
determinations.
\\We first sought for long term correlations in the time series of Figure~\ref{fig:1} using spectral and detrended fluctuation
analysis (see~\ref{sec:Spectral}). Figure~\ref{fig:1}E shows the
power spectrum $S(f)$ variations with the frequency $f$ on a log-log
scale. First, we note that the power spectrum has a broadband
characteristic, typical of stochastic and chaotic signals.
Furthermore, for the three statistics tested, the power spectrum
scales as a power-law of the frequency, for frequencies $f
\gtrapprox 2 \times 10^{-9} \quad instruction^{-1}$ (i.e. for
periods lower than the major period of repetition) with spectral
exponent $\beta \approx 1.3$. Detrended fluctuation analysis for the
three time series is presented Figure~\ref{fig:1}F. Here again, for
time scales lower than the major period of repetition, we observe
for the three time series a power-law relationship between $F(n)$
and $n$, with an exponent $\alpha \approx 1.13$.
\begin{figure*}[!htb]
  \includegraphics[scale=0.7]{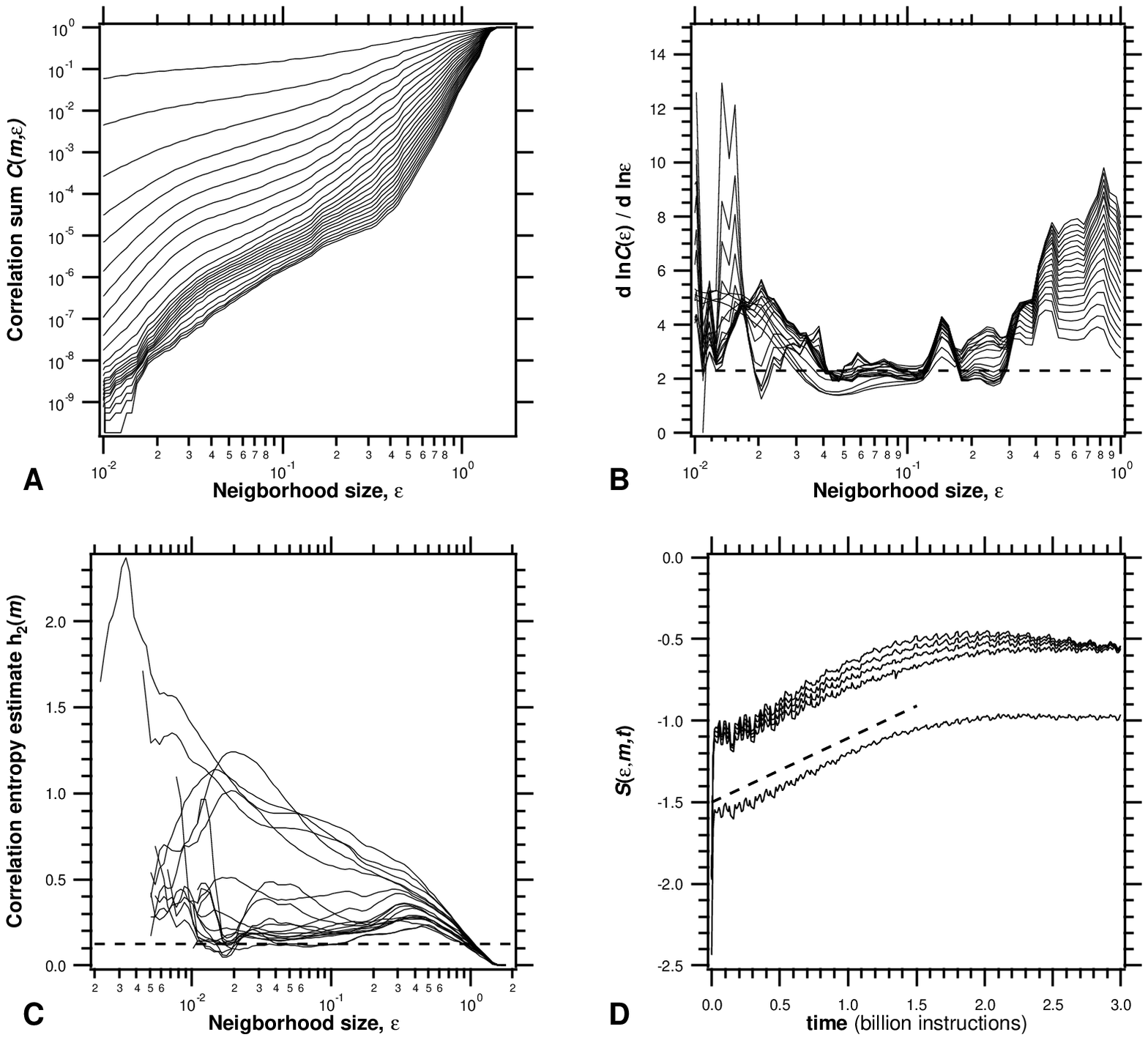}
  \caption{\label{fig:3} Characterization of the attractor for bzip2 after reconstruction by embedding of the IPC trace.
  (A) Log-log plot of the correlation sums $C(m,\epsilon)$ as a function of the considered neighborhood $\epsilon$, with $m$ ranging,
  from top to bottom, from $1$ to $25$ (with increment of 1). (B)
Corresponding local slopes $d \ln C(m,\epsilon)/d \ln \epsilon$ with
$m$ ranging, from top to bottom, from $25$ to $7$ (with decrement of
1). The dashed line indicates the value estimated for the
correlation dimension $D_2 =2.3 \pm 0.3$ (C) Corresponding estimates
of the correlation entropy $h_2$. The dashed line indicates the
value 0.125, yielding $h_2 \approx 1.2$ bits/billion instructions.
(D) Estimation of the largest Lyapunov exponent (see
~\ref{sec:lambdamax}) using the R\"{o}senstein's (bottom trace) or
Kantz's (top traces) method with $m=12$ to $15$ (from top to
bottom). The dashed line indicates a power-law growth with exponent
0.42, yielding an estimate of $\lambda_{max}\approx 0.60$
bits/billion instructions. Embedding parameters are those of
figure~\ref{fig:2}.}
\end{figure*}
 Note that the two
independently-obtained exponent values satisfy the relationship
$\alpha = (1+\beta)/2$~\cite{Buldyrev1995, Heneghan2000}, which is
an indication
 of the consistency of these values~\cite{Rangarajan2000}.
\\These results first show that \verb"bzip2" performance statistics
display $1/f^\beta$-noise. This reveals the absence of a
characteristic time scale for the duration and recurrence of the
performance variations (at least for those variations with
time-scales shorter than the major period of repetition). Hence
\verb"bzip2" performance time series display a high level of
self-similarity. Furthermore, the value obtained for $\alpha$ is
greater than 0.5 (and $\beta>1$). This is a sign of the existence of
persistent long-range correlations inside the time series i.e. a
large (compared to the average) IPC or cache miss rate value is more
likely to be followed by a large IPC or cache miss rate value and
vice versa. The presence of these correlations is a first argument
to exclude the
 possibility of (noncorrelated) noise as a source of variability of the
 traces.\\
 To study further the dynamics, we reconstructed its attractor through embedding of the IPC time series. The embedding parameters (delay
 $d$ and dimensions $m$ see~\ref{sec:Embedding}) were estimated to $d=153$ million instructions and
$m=14$. Figure~\ref{fig:2}A presents the thresholded recurrence
plot. We first note that the two consecutive phases displayed by
\verb"bzip2" (see Figure~\ref{fig:1}A) are clearly recognizable from
the recurrence plot, indicating that their recurrence rates may be
significantly different. Interestingly, the plot presents many
interrupted diagonal lines beside single isolated points.
Furthermore, these lines exhibit some level of periodicity, which
could be a sign that the system contains unstable periodic orbits
(UPOs)~\cite{Bradley2002}. This kind of structure is typical of
chaotic systems~\cite{Eckmann1987}. We also present in
Figure~\ref{fig:2}B the first-return map of the Poincar\'e section
at IPC minima of the reconstructed attractor. The map is highly
structured, with several mono-dimensional parts, which is another
sign of low dimensional chaotic dynamics.
\\Thus, these first elements plead in favor of a
chaotic component in \verb"bzip2" performance time series. Chaotic
dynamics being a manifestation of nonlinear systems, we next sought
the presence of nonlinearities in this time series using surrogates
data (see~\ref{sec:surr}). Figure~\ref{fig:2}C shows a segment of
the time series (upper trace) together with the corresponding
surrogate (lower trace). Visual comparison of these two signals
already suggests their dissimilarity. To confirm visual inspection,
we performed statistical tests, quantifying nonlinearity with two
different statistics. The null hypothesis was that the IPC trace
could be generated by a linear, possibly rescaled, Gaussian random
process. Both quantification statistics yield to reject the null
hypothesis at the 95\% level of significance, hence confirming the
nonlinear nature of the IPC execution trace.\\
To study the reconstructed attractor in more details, we next
characterized its geometrical properties. Figure~\ref{fig:3}A
displays a log-log plot of the correlation sums $C(m,\epsilon)$
obtained for various dimensions $m$, versus the neighborhood size
$\epsilon$. A power-law regime between $\epsilon \approx 0.02$ and
$\epsilon \approx 0.3$ is apparent for high $m$ values. Furthermore,
the corresponding slopes in this regime (the exponents of the
power-laws) seem to tend to a rather constant value at high $m$.
This scaling is confirmed in Figure~\ref{fig:3}B that shows the
local slopes $d \ln C(m,\epsilon)/d \ln \epsilon$ of the curves of
Figure~\ref{fig:3}A. For $0.03 \lessapprox \epsilon \lessapprox 0.3$
and $m \gtrapprox 9$, the local slopes collapse to a $m$- and
$\epsilon$-independent value of $\approx 2.3$. The occurrence of
such a scaling regime is a strong sign that the observed variability
in the dynamics is not caused by a random source, thus confirming
the hypothesis of a chaotic behavior. The value in the scaling
regime is an estimation of the correlation dimension of the
attractor, $D_2 = 2.3 \pm 0.3$. The correlation dimension is one
measure of the attractor fractality. Thus, its non-integer value
might be an indication that the attractor for \verb"bzip2"
performance dynamics is a fractal object, like most of the chaotic
strange attractors. However, as is very often the case with
real-life systems,  our estimation of $D_2$ is not precise enough to
exclude an integer value, so that the attractor fractality cannot be
asserted in the light of our present results. However, the (low)
value of $D_2$ remains a strong indication the \verb"bzip2"
performance displays low-dimensional chaos. \\
\begin{figure*}[!htb]
  \includegraphics[scale=0.7]{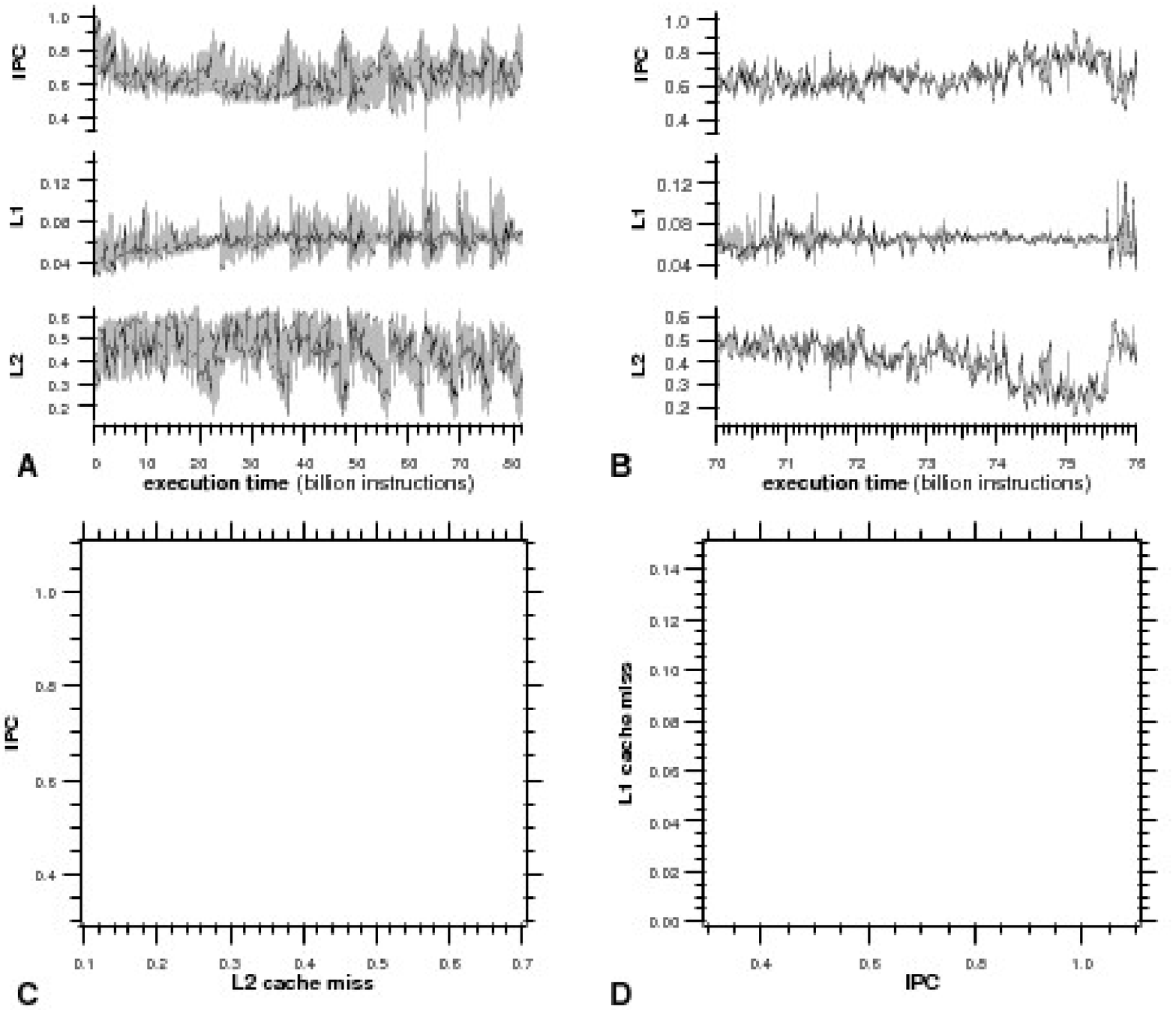}
  \caption{\label{fig:4} Execution traces for the program vpr. (A) Evolution of the number of instructions executed per cycle (IPC), the rate of L1 cache misses (L1)
  and that of L2 cache misses (L2) over the complete program execution and (B) enlargement of the segment comprised between
  70 and 76 billion instructions. Also shown are projections of the
dynamics attractor (C) in the IPC-L2 cache miss rate phase
  plan and (D) in the L1-IPC phase plan.}
\end{figure*}
The correlation sums can also be used to estimate the corresponding
correlation entropy $h_2$. Figure~\ref{fig:3}C presents the
resulting estimates as a function of $\epsilon$ and for $m$ varying
from 7 to 25. The value of $h_2$ can be estimated in the scaling
regime observed in Figure~\ref{fig:3}B, but must be extrapolated at
large $m$. Accordingly, our estimate on the basis of
Figure~\ref{fig:3}C (dashed line) yields $h_2 \approx 1.2$
bits/billion instructions.
\\A very strong indication of chaotic dynamics is
sensitivity to initial conditions (SCI). To quantify SCI in our
systems, we tried to estimate the largest Lyapunov exponent from our
reconstructed attractor (Figure~\ref{fig:3}C) using both
Kantz's~\cite{Kantz1994} (top four curves) and
R\"{o}senstein's\cite{Rosenstein1993} (bottom curve) methods.  The
occurrence of a positive Lyapunov exponent is amongst the strongest
indications of chaotic dynamics. Both methods result in similar
curve shapes. Although the data are far from ideal, a linear part at
short times can be distinguished in all these curves. The slope of
these linear parts provides us with an estimate for the largest
Lyapunov exponent $\lambda_{max}\approx 0.60$ bits/billion
instructions. Alternatively, the largest Lyapunov exponent can also
be measured from the Poincar\'{e} map. Our estimations on the basis
of Figure~\ref{fig:2}B (data not shown) yield a somewhat higher, but
comparable estimate ($\lambda_{max}\approx 1.22$ bits/billion
instructions). These estimates can be compared to the correlation
entropy $h_2$, which is a lower bound of the sum of all the positive
Lyapunov exponents of the system (see~\ref{sec:Correl}). Hence our
estimates for $h_2$ and $\lambda_{max}$ are readily comparable,
thereby further supporting the consistency of our measurements.
\\The measurements and analysis presented so far were essentially
obtained on the basis of a reconstruction of the attractor using the
IPC time series. We also carried out most of these analyzes using
the other two time series (L2 and L1 cache miss rates) for attractor
reconstruction and varied the averaging window $\tau_{av}$ ($\tau_{av}=10^6,\,10^7\, \mathrm{or}\,10^8$ instructions, see~\ref{sec:trace}). All these conditions yielded comparable
values and confirmed that \verb"bzip2" performance dynamics display
low dimensional deterministic chaos. Furthermore, we analyzed
\verb"bzip2" performance dynamics with another data input (namely,
the \verb"program" input of the \verb"ref" data set, see
~\ref{sec:trace}). Although these dynamics displayed possibly lower
SCI ($\lambda_{max} \approx 0.5$ bits/billion instructions), all
tested indicators confirmed the presence of chaotic dynamics,
indicating that their origin is more probably rooted into the
program/architecture interaction than to be found in a
data-dependent mechanism.
\begin{figure*}
  \includegraphics[scale=0.7]{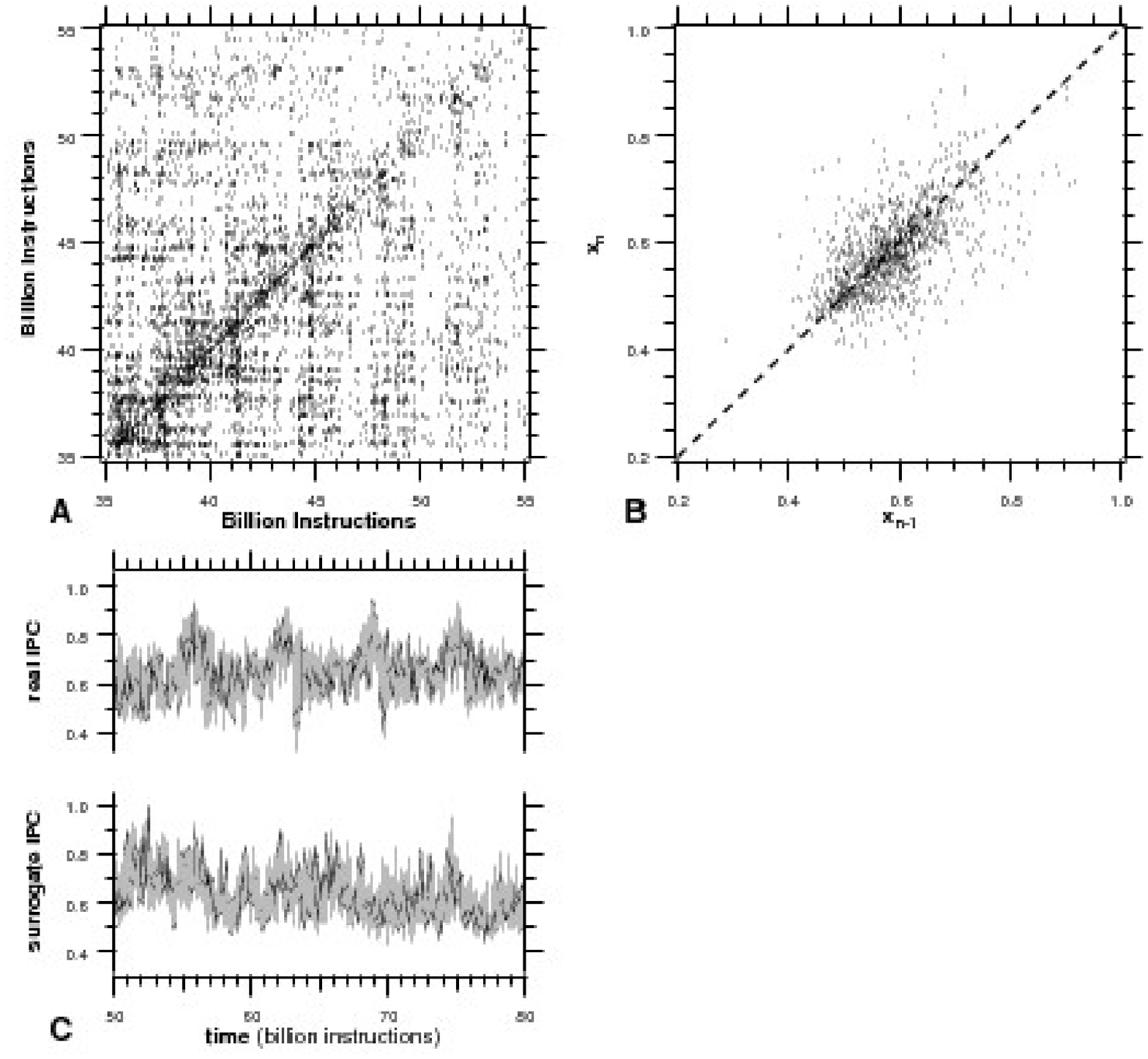}
  \caption{\label{fig:5} Study of the IPC trace for vpr after attractor reconstruction by embedding. (A) Recurrence plot thresholded at
  0.079. Note that, for readability, we only present the data obtained for the segment ranging from 35 to 55 billion instructions.
  (B) First return map for the Poincar\'e section at IPC minima. The dashed line indicates the
  first diagonal. (C) Surrogates data developed from the complete IPC trace. Shown are enlargements of the part comprised between
  50 and 80 billion instructions of the original IPC time series (upper
trace) and corresponding surrogate trace (lower trace). Embedding
parameters: delay
  $d=350$ million instructions, dimension $m=5$.}
\end{figure*}
\\The magnitude of the largest Lyapunov exponent
quantifies the attractor's dynamics in information theoretic terms.
As a crude interpretation, it measures the rate at which the system destroys information. For
instance, suppose one knows the number of instruction executed per
cycle for \verb"bzip2" at some initial time $t_0$ with good
accuracy, say $0.01\%$ (13 bits). Because of the intrinsic
sensitivity to initial conditions (say, in average,
$\lambda_{max}\approx 0.9$ bits/$10^9$ instructions), 0.9 bits of
this information will be lost, in average, every billion
instructions. In other words, after 15 billion instructions (i.e.
$\approx 1/8$ of the total program execution length), the IPC number
would be no more predictable. Note however that the magnitudes of the Lyapunov exponents quantify \textit{average} convergence or divergence rates (over the phase space), but in fact, the degree of predictability can vary considerably throughout phase space~\cite{Abarbanel1991}. Hence it is possible to loose predictability exponentially fast in some part of the dynamics, while regaining it later on.\\
To compare with other chaotic systems, these
values must be related to the duration of an average orbit around
the attractor, which is $\approx 430$ million instructions, yielding
a value ranging from 0.26 to 0.52 bits/average orbit. Although lower
than that of the Lorenz system ($\lambda_{max} = 1.36$ bits/orbits),
this value is comparable to that obtained for the R\"{o}ssler system
($\lambda_{max} = 0.78$ bits/orbits)~\cite{Wolf1985}, a classical
model for deterministic chaos.
\\Finally, we note that this kind of dynamics is not restricted to \verb"bzip2". Amongst the tested Spec benchmarks, we evidenced
deterministic chaos with other programs including \verb"galgel" and
\verb"fma3d", and obtained some indications of it (albeit not
conclusively) for \verb"gzip" and \verb"ammp".
\begin{figure*}
  \includegraphics[scale=0.7]{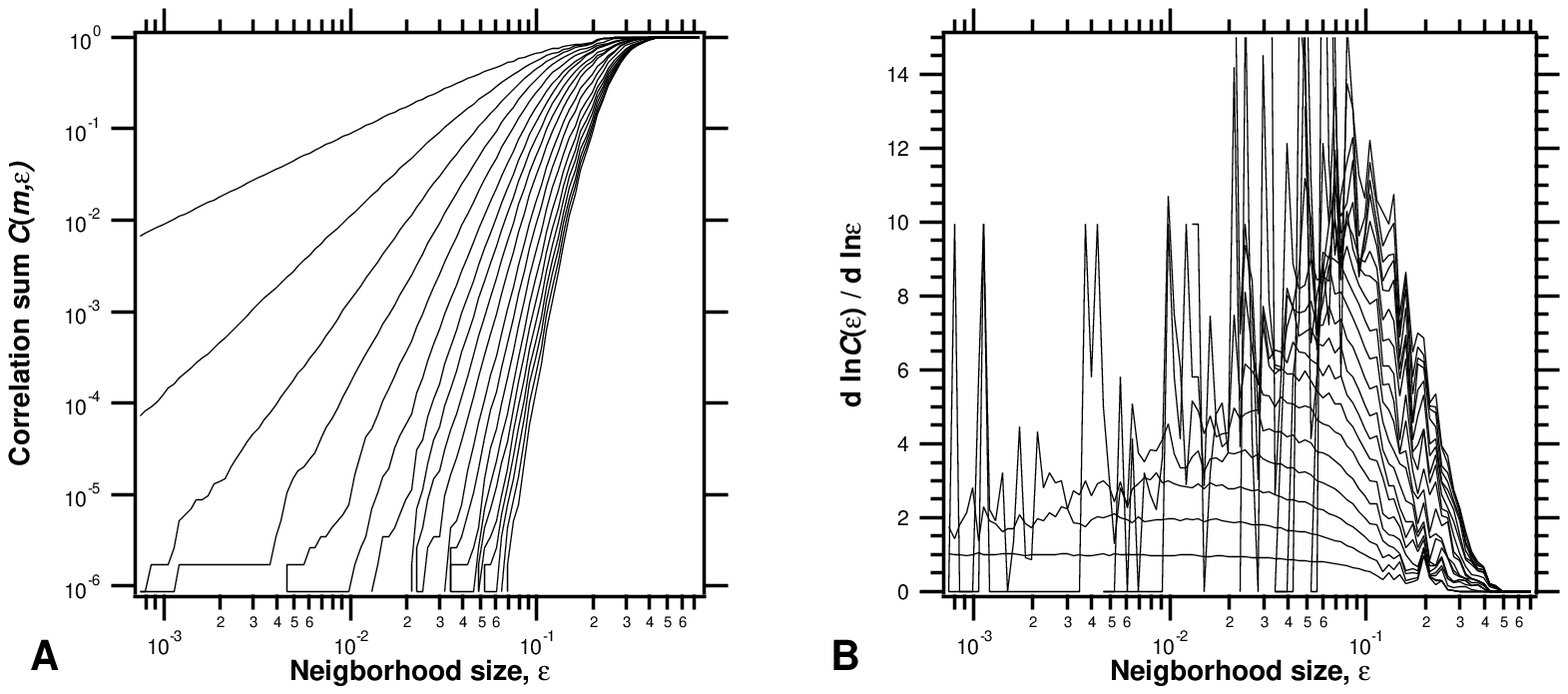}
  \caption{\label{fig:6} Characterization of the attractor for vpr after reconstruction by embedding of the IPC trace.
  (A) Log-log plot of the correlation sums $C(m,\epsilon)$ as a function of the considered neighborhood $\epsilon$, with $m$ ranging,
  from top to bottom, from $1$ to $20$ (with increment of 1). (B)
Corresponding local slopes $d \ln C(m,\epsilon)/d \ln \epsilon$ with $m$
ranging, from top to bottom, from $20$ to $1$ (with increment of 1).
Embedding parameters are those of figure~\ref{fig:5}.}
\end{figure*}
\subsection{\label{sec:vpr}Second example: vpr time series}
Evolution of the three studied performance statistics for the
program \verb"vpr" are shown Figure~\ref{fig:4}. As compared to
\verb"bzip2", the dynamics are much more variable and lack real
regular behaviors. Likewise, the projections onto phase plans
display clouds of points lacking clear inner structures. We
reconstructed the attractor of the dynamics through embedding of the
IPC time series (with $d=350$ million instructions and $m=5$).\\
Figure~\ref{fig:5}A presents the thresholded recurrence plot for
this embedding. In opposition to the recurrence plot obtained for
\verb"bzip2" (Figure~\ref{fig:2}A), vpr recurrence plot only
displays isolated points (no diagonal lines) that are much more
homogeneously distributed (distribution structures are not easily
visible). Likewise, the Poincar\'{e} map presented
Figure~\ref{fig:5}B displays a rather homogeneous scattering of the
points over the first diagonal. The aspect of these two figures are
first indications that vpr variability is neither periodic, nor the
result of chaotic dynamics. In agreement with these conclusions, we
note that, even if the corresponding surrogates
(Figure~\ref{fig:5}C) are visually similar to the original IPC time
series, statistical tests for the presence of nonlinearities in vpr
performance dynamics could not decide between the presence or the
absence of nonlinearity in the original trace. This can be
considered as a first indication that, while not chaotic nor
periodic, this time series might neither result from a really
stochastic process.
\begin{figure*}
  \includegraphics[scale=0.8]{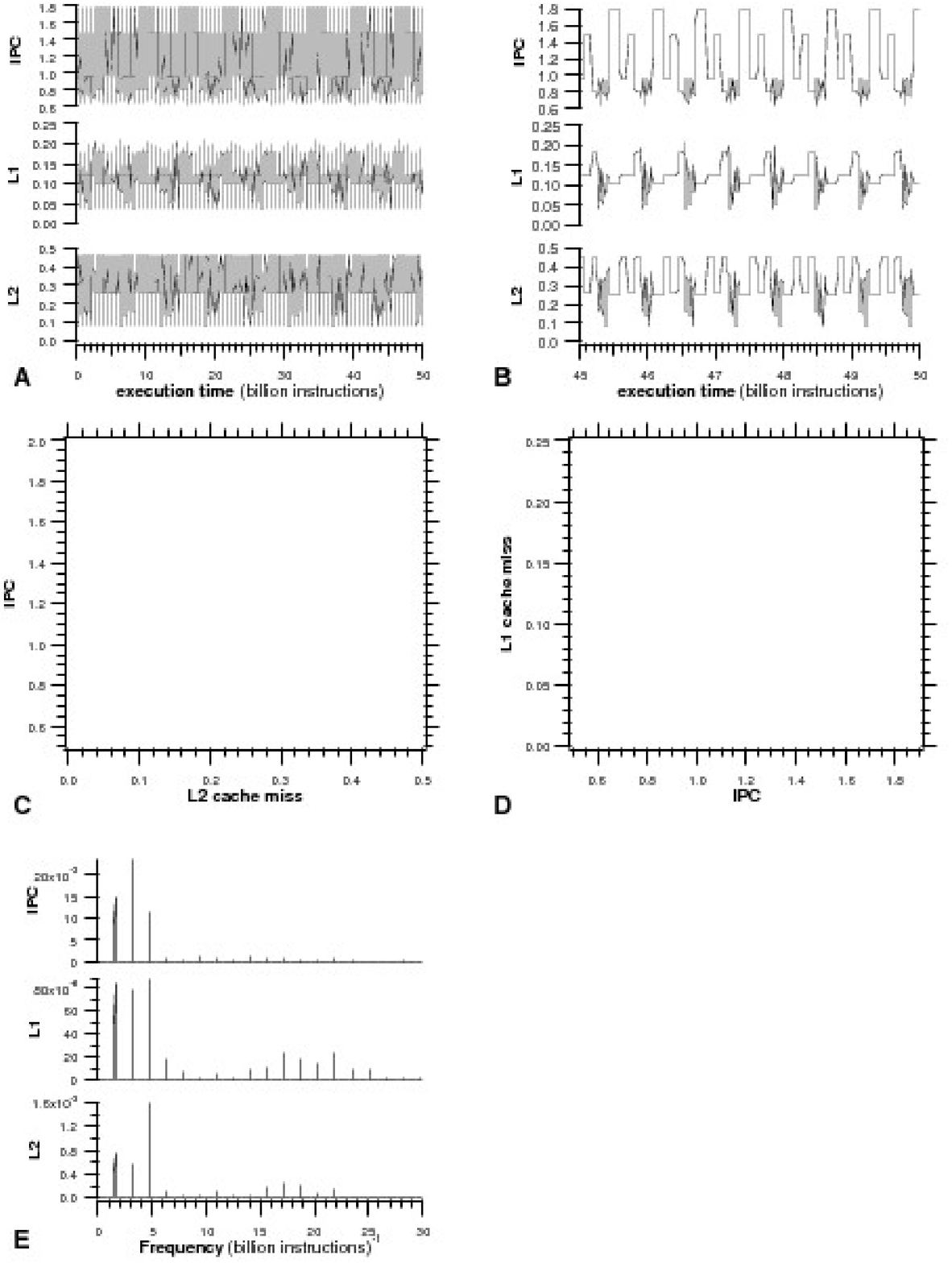}
  \caption{\label{fig:7} Execution traces for the program applu. (A) Evolution of the number of instructions executed per cycle (IPC),
  the rate of L1 cache misses (L1)
  and that of L2 cache misses (L2) over the first 50 billion instructions (the entire program execution consists of circa
  220 billion instructions) and (B) enlargement of the segment comprised between 45
  and 50 billion instructions. Also shown are projections of the dynamics attractor (C) in the
IPC-L2 cache miss rate phase
  plan and (D) in the L1-IPC phase plan. The power spectra for the three time series are presented in (E)}
\end{figure*}
\\Figure~\ref{fig:6} shows the corresponding correlation sums for
$m$ ranging (from top to bottom) from 1 to 20. Although a regime
with power-law behavior is observed for each curve, the slopes of
the corresponding linear parts do not seem to saturate to a constant
value with increasing $m$. This is confirmed by examination of the
respective local slopes presented Figure~\ref{fig:6}B. In opposition
to the corresponding plot for \verb"bzip2" (Figure~\ref{fig:3}B),
this figure fails to show any scaling regime, whatever the
$\epsilon$- or $m$-range considered. Absence of saturation of the
correlation sum exponents at high $m$ is another indication that,
contrarily to \verb"bzip2", the high variability and irregularity of
\verb"vpr" performance dynamics are not imputable to chaotic
dynamics, but result from some ``high dimensional'' non chaotic
process.
\\Amongst the Spec benchmarks we inspected, a similar behavior was also observed for \verb"art", and
suspected for several other programs, such as
\verb"crafty", and (albeit to a lesser extend) \verb"ammp",
\verb"gcc", or \verb"gzip".
\subsection{\label{sec:applu}Third example: applu time series}
Our last example concerns the program \verb"applu", a scientific
computing application. A simple inspection of the time series is
enough to evidence the regularity of the three performance
statistics (Figure~\ref{fig:7}A and B). Projections in the phase
plans (Figure~\ref{fig:7}C and D) provide a striking representation
of a multiply folded one-dimensional attractor, reminiscent of
multi-dimensional limit cycles. These periodic oscillations are so
regular that the folded attractors display an almost null noise
level. In agreement with these observations the power spectra for
the three statistics (Figure~\ref{fig:7}E) are typical of periodic
patterns, with a major frequency ($f\approx 1.6\times 10^{-9}$
instructions corresponding to a period of $\approx 0.6$ billion
instructions, compare with Figure~\ref{fig:7}B) and its harmonics
dominating the spectrum.
\\Taken together, these results unambiguously show the existence of
programs with highly regular performance traces. Besides
\verb"applu", such a behavior was also evidenced for other Spec
benchmark programs such as \verb"apsi".

\section{\label{sec:Conclusion}Discussion}
\subsection{Potential sources of seemingly stochastic dynamics}
An intriguing result of this paper is that the performance traces of
several program are not periodic nor chaotic, but display a high
level of aperiodic fluctuations (such as \verb"vpr"), that appear
similar to stochastic dynamics from the point of view of the
nonlinear methods we used. This may sound counterintuitive because
the underlying microprocessor operations are deterministic by
nature. Several sources of aperiodic variability in the
performances can be evoked.\\
First, a potential source of aperiodicity resides in the simulated
programs themselves. A great number of the programs from the SPEC
benchmark are scientific codes and many of them use
pseudo-random numbers. Albeit pseudo-random number
generators are also purely deterministic routines, their output is
hardly distinguishable from truly random numbers. This could in part
be implied in the apparently stochastic behaviors we observed.
Second, one must not forget that the metrics we studied are indirect
measurements of the microprocessor state. In other words, while the
microprocessor deterministically processes the program flow, we only
record its performance. It has recently been remarked that the
correlation between the code being executed and the performance can
vary widely~\cite{Annavaram2004}. In other words, for some programs,
performance metrics are highly dependent on the execution history,
so that two executions of the same code piece during a single
program can have performance metrics that vary considerably. This
source of variability could also in part explain the behavior of
``high dimensional'' traces such as \verb"vpr".\\
Furthermore, recall that to distinguish between chaotic and
stochastic signals, nonlinear time series methods usually make use
of the fact that, contrarily to stochastic dynamics, chaotic ones
are ``bounded'' (their attractor have a finite dimension). In the
same way that these methods could not distinguish purely random
numbers from pseudo-random numbers generated by modern libraries,
the \verb"vpr" traces could abusively appear stochastic to them. In
fact, even simple deterministic processes can yield behaviors that
appear stochastic to visual inspections (see for example Chapter 4
in~\cite{Wolfram2002}). Incidentally, we note that the IPC time
series of \verb"vpr" is strikingly similar to the apparently
stochastic fluctuations of the simple deterministic recursive
iteration presented page 130 (bottom trace) in~\cite{Wolfram2002}.
Hence, what can rigorously be said of the \verb"vpr" case is that it
is highly fluctuating, and that these fluctuations are neither
regular (periodic) nor chaotic, but result of a ``high dimensional''
process.
\subsection{Chaotic performance time series and predictability}
The other specific conclusion drawn by this study is that the high
variability in the time-evolution of the performances during the
execution of several programs can be imputed to deterministic chaos.
This result seems important because it implies that performance
predictability based on short sampled sequences might be impossible
and because in a more general perspective, it reveals the high
intricacy of the processes determining instantaneous microprocessor
performances. However, its interpretation must be handled with great
care. First, the obtained results apply to instantaneous
performances \textit{only} and do not imply \textit{other} aspects
of microprocessor operations. For instance, they neither imply that
program execution itself (i.e. the instruction flow handled by the
processor) is chaotic or unpredictable. In particular, they do not
imply that the program final result might be variable nor
unpredictable.\\
Chaotic dynamics are known to occur in systems where the variables
are in great number and/or interact through nonlinear relationships.
Modern microprocessors include a large number of hardware mechanisms
that are dedicated to improve performance (speculative execution,
branch predictors, prefetchers, memory and instruction caches,
pipelines...). As a result, the precise number of cycles needed to
execute a given instruction sequence depends on a huge number of
internal states of hardware components. For instance, the precise
number of clock cycles needed to execute a simple instruction
sequence including at least one conditional branch and one
load/store instruction depends, among others, on the state of the
branch predictor mechanism (which is usually history-dependent)
corresponding to this branch, on the states of the different caches
of the memory hierarchy (presence or absence of the data), the
precise state of all instructions in all stages of the execution
pipeline and in the numerous buffers included in the processors.
Furthermore, these different internal states are usually related
through nonlinear relationships (for instance, a branch prediction
error can lead to a complete flush of the execution
pipeline, which may, in turn modify this branch predictor state).\\
Hence, exact knowledge of the state of the set of
performance-determining mechanisms at a given time is unattainable.
This property is so strong that it has recently been used to build
powerful pseudo-random number generators based on the
unpredictability of the internal microprocessor
states~\cite{Seznec2003}. As a result, two states of the
performance-determining mechanisms that appear arbitrarily close
with respect to the partial information possessed by the observer,
can in fact be different. Because performance critically depends on
the \textit{global} state, the performance evolutions starting from
these two seemingly similar states can be highly different. This
might account for the observed sensitivity to initial conditions
(i.e. chaos). Note however that further work is needed to understand
why these properties manifest during the execution of certain
programs only, while it seems not to be prominent for others.
\subsection{Relevance to practical applications}
Finally our results may have some practical importance in the field
of performance modeling. To predict the effect of a given hardware
mechanism, computer architects use detailed simulations of the
microprocessor performance during program execution. Because these
detailed simulations are highly demanding on calculation time,
several methods have been developed to estimate the average
performance on the basis of a subsample of the entire execution
trace. Our result that several program traces (such as vpr) display
dynamics that are closed to stochastic ones could be useful in this
framework. Indeed, this usually means that the obtained surrogates
data are very similar to the corresponding real traces (see figure
5C, for instance). Hence, for these programs, it is possible to
consider generating long surrogates data (at very low computational
costs) from a short sample of the real trace, and use these
synthetic traces to estimate the average metric (average ipc, for
example) during a real
execution of the program.\\
Conversely, our results indicate that for those programs endowed
with chaotic behaviors (such as bzip2 or galgel), it might be very
delicate to predict the actual evolution of the considered
performance metric on the basis of extrapolations from a short
sequence of the real trace. Hence, for these programs, our results
suggest that an efficient strategy for predicting the actual average
value of the metric under consideration on the ground of a sample of
its real trace would be to base the estimation on several samples
extracted from the real trace, even in a random way. Actually this
method is used by one of the most powerful tool developed for
performance prediction~\cite{SMARTS}. Yet, it should be recalled
that variations on a strange attractor are bounded so that the
existence of these difficulties does not exclude the possibility to
predict accurate \emph{average} values, which is the aim of most of
these methods~\cite{SIMPOINT,simpoints}. Finally, the necessity to
adapt the performance simulation/sampling technique as a function of
the program under consideration has recently been pointed
out~\cite{Annavaram2004}. We think our results might account for a
rationale of this necessity.

\newcommand{\noop}[1]{}


\end{document}